\documentclass[11pt]{article}   	% use "amsart" instead of "article" for AMSLaTeX format
\usepackage{geometry}                		% See geometry.pdf to learn the layout options. There are lots.
\usepackage{graphicx}				% Use pdf, png, jpg, or eps§ with pdflatex; use eps in DVI mode
\usepackage{amsmath,amssymb}
\usepackage{hyperref}

\newcommand{\NP}{\textsf{NP}}
\newcommand{\QIP}{\textsf{QIP}}
\newcommand{\QMA}{\textsf{QMA}}
\newcommand{\PSPACE}{\textsf{PSPACE}}

\title{Synergies Between Operations Research and Quantum Information Science}
\author{Ojas Parekh\\ Quantum Algorithms and Applications Collaboratory (QuAAC) \\ Sandia National Laboratories\\ \texttt{odparek@sandia.gov}}
\date{}							% Activate to display a given date or no date

\begin{document}

\maketitle

\abstract{This article highlights synergies between quantum information science (QIS) and operations research for QIS-curious operations researchers (and vice-versa).}

\section{Introduction}

Operations researchers are no strangers to transferring their expertise to new domains to realize an impact.  Such endeavors often entail understanding enough domain specifics to effectively build models and solve problems.  This can be an iterative and challenging process, as sometimes idiosyncrasies that have been internalized by domain experts need to be sussed out.  However, overcoming such obstacles may become particular points of pride, in addition to overall success.  Even though quantum information science (QIS) is a broad, vibrant, and intensely growing field, I advocate approaching QIS the same way we might a more specialized domain.  Instead of being daunted by, for example, never having taken a quantum physics course, we might try to stick to mathematical descriptions or other abstractions, with the understanding of likely being oblivious to a body of underlying intuition that has been well earned by physicists.  Time and experience may help remedy the latter if desired.  

I highlight seminal or recent advances in QIS attained through the lens of optimization.  I also offer suggestions on how engaging with QIS might lead to advances in more traditional operations research (OR).  Finally, I present strategies for operations researchers to engage QIS.  I encourage operations researchers to cultivate new synergies with QIS.
         
\section{Quantum Information Science applications of Operations Research}         

\paragraph{\bf Properties of quantum states and channels.} Without worrying about additional details or quantum-mechanical interpretations, we may think of a \emph{quantum state}, $\rho$, on $n$ \emph{quantum bits} (qubits) as a $2^n \times 2^n$ matrix with complex entries.  Many basic properties of quantum states and the quantum \emph{channels} describing operations on them may be readily cast as semidefinite programs.  In fact the defining properties of a state $\rho$ are that it has trace equal to one and is positive semidefinite.  For more details, a self-contained account of five accessible applications of semidefinite programming to properties of quantum states and channels appears in~\cite{siddhu2021five}.  

The high-level perspective of the remainder of the article will not expect an understanding of qubits or quantum states, beyond the fact that quantum states are exponentially large in the number of qubits.  The latter opens the door to potential exponential advantages over classical computation, as clever physical manipulations of $n$ qubits may enable nature to implicitly process exponentially large quantum states in meaningful and useful ways.  However, the same kind of statement could be made about randomized classical algorithms, where manipulating $n$ random bits yields an implicit distribution over an exponentially large set of outcomes.  Thus we seek to identify features of quantum physics that are not accessible classically.

\subsection{Differentiating classical and quantum physics} 
What does a universe endowed with quantum physics offer that is simply impossible under the laws of \emph{classical} physics (i.e., conventional non-quantum physics)?  Let me highlight such an example.  In a \emph{nonlocal game} Alice and Bob are to be posed questions $x$ and $y$ and must agree on a strategy that generates answers $a$ and $b$.  The value of a nonlocal game, $V(a,b\mid x,y) \in \{0,1\}$ determines correctness of the answers, and Alice and Bob's goal is to maximize the expectation of $V$ over a given distribution on $x,y$ (and potential randomness in selecting $a,b$).  It turns out that random strategies do not offer any advantage over deterministic strategies; however, strategies where Alice and Bob each have one of a pair of \emph{entangled}\footnote{Rather than offering a concise but potentially misleading description of entanglement here, I suggest the popular article, \cite{wilczek2016entanglement}.} quantum bits (qubits) are able to outperform classical strategies (see the survey, \cite{brunner2014bell}).  

Advantages offered by quantum strategies to nonlocal games are intimately related to John Bell's seminal tests for nonlocality (or ``quantumness'') in physical systems.  Quantum violations of Bell inequalities, that classical physical systems must satisfy, have been demonstrated experimentally under a variety of settings (\cite{brunner2014bell}, Section VII).  The values of nonlocal games may in turn be approximated by hierarchies of semidefinite programs (SDPs) (\cite{brunner2014bell}, Section II.C); however, the size of the resulting programs grows exponentially (or worse) in the size of the game.  Imagine the gratification in finally successfully solving or analyzing a thorny SDP and subsequently receiving a message from your experimental-physicist friend that nature agrees with your findings -- a nice pat on the back from the universe. Nonlocal games have fundamental connections to models of (quantum) computation \cite{mousavi2021nonlocal}, and a nonlocal games perspective has recently enabled a landmark result resolving two longstanding problems: Tsirelson's problem in quantum mechanics and Connes' embedding problem in operator algebras \cite{ji2021mip,palazuelos2016survey}.

\subsection{Computational quantum advantages}
A foremost direction in QIS is leveraging non-classical features of quantum physics to realize forms of \emph{quantum computation} that are able to enjoy advantages over conventional classical computation.  Shor's seminal quantum algorithm for factoring integers runs exponentially faster than any known classical algorithm (see e.g., \cite{nielsen2010quantum}, Section 5.3); however, as far as we currently know, exponentially faster classical factoring algorithms might well exist.  Yet, if we consider computational resources beyond execution time, exponential quantum advantages over \emph{best-possible} classical algorithms are known.  

One such setting is query complexity, where we are only concerned with the number of queries to a problem's input data rather than overall execution time.  Exponential quantum advantages in query complexity were among the first quantum algorithms discovered (see e.g., \cite{nielsen2010quantum}, Section 1.4).  The scope for exponential advantages in query complexity for graph problems is generally well understood (\cite{ben2020symmetries}; see the related survey, \cite{montanaro2016survey}).  Remarkably, finding the best quantum algorithm for a problem in the query model can be captured, within constant factors, as SDPs \cite{reichardt2011reflections,barnum2003quantum}, and this perspective has helped design quantum graph algorithms (e.g., \cite{delorenzo2019applications}).   Recent work has characterized the precise query complexity in terms of the completely bounded norm of a tensor \cite{arunachalam2019quantum}, which in turn is expressible as an SDP \cite{gribling2019semidefinite}.  A relaxed notion of query complexity in expectation turns out to be intimately related to the Sherali-Adams hierarchy in the classical case and the Lasserre/Sum-of-Squares (SoS) hierarchy in the quantum case \cite{kaniewski2015query}.  For open problems in quantum query complexity see \cite{aaronson2021open}, and for a broader survey of exponential quantum speedups see \cite{aaronson2022much}.

\subsection{Building better quantum computers}
The biggest open question in quantum computing is perhaps whether we can indeed design and engineer scalable fault-tolerant quantum computers to realize theoretically supported quantum advantages.  The world is a particularly hostile place for a quantum computer, with magnetic fields, variations in temperature, and a host of other sources of \emph{noise} that are disruptive to computation.  Noise induces errors that are amplified the longer a computation executes.  Schemes to correct such errors are known; however, they demand considerable overhead in terms of extra error-correction qubits.  Designing efficient quantum error correction schemes with desirable resiliency properties may be cast as an optimization problem, and semidefinite programming techniques have been used to design and analyze such schemes \cite{fletcher2007optimum,kosut2009quantum,berta2021semidefinite}.          

Quantum processors may be built upon different physical substrates, though the selection is drastically constrained compared to classical processors.  Each brings unique design and engineering challenges, as well as associated optimization problems.  Quantum processors are generally put through quantum characterization, validation and verification hoops to ensure they behave as expected (see the tutorial, \cite{kliesch2021theory} and review, \cite{eisert2020quantum}).  Well-behaved quantum processors sit under software stacks that offer further opportunities for optimization, including compiling high-level quantum algorithms into quantum circuits consisting of native gates.  Interdisciplinary teams including operations researchers are increasingly addressing optimization problems at many levels of quantum computer system design \cite{nannicini2022optimal,nagarajan2021quantumcircuitopt,tang2021cutqc,fei2022binary,morvan2022optimizing,banchi2020convex}.

\subsection{Relaxations for problems in quantum physics}
The above applications suggest a recurring theme: convex programs naturally model quantum-mechanical phenomena, but faithful models require exponential-size programs in the number of qubits.  A technique OR may bring to the table is finding smaller but reasonably strong relaxations of such exponential-sized convex programs.  This could help accelerate approaches for exactly or approximately computing quantities of physical interest.  I will illustrate this concept by drawing connections between discrete optimization problems and quantum counterparts, using the well-known classical Max Cut problem as an example.

\paragraph{\bf Max Cut as an eigenvalue problem.}
For a graph $G = (V,E)$ on $n$ vertices, Max Cut seeks to find a set $S \subseteq V$ that maximizes the number of edges between $S$ and $V\setminus S$.  We will encode Max Cut as finding the largest eigenvalue of a matrix exponentially large in $n$, for more direct comparison with problems from quantum physics.  Imagine the columns and rows of a diagonal matrix $H \in \mathbb{R}^{2^n \times 2^n}$ are labeled with the $2^n$ possible vertex sets $S \subseteq V$, and set $H_{S,S}$ to the number of edges in the cut $(S, V\setminus S)$.  Now since $H$ is diagonal, $\lambda_{\max}(H)$ (the maximum eigenvalue of $H$) is the maximum cut value, achieved by a set $S^*$.  The corresponding eigenvector $v^* \in \mathbb{R}^{2^n}$ corresponds to a basis vector with a one in the position labeled $S^*$ and zeros elsewhere.  Even though $v^*$ lives in an exponentially large space, it has a succinct description that only depends on $S^*$.  The matrix $H$ itself also has a succinct description that only depends on the edges in $G$.  In summary, Max Cut (or discrete optimization problems in general\footnote{$H$ could be labeled with subsets of any discrete domain, with infeasibility modeled by large-magnitude values.}) may be cast as finding $\lambda_{\max}(H)$ for a diagonal matrix $H$ that is exponentially large in $n$ with a description of size polynomial in $n$.  

\paragraph{\bf Sampling-based problem models.} What happens if a symmetric $H$ is not required to be diagonal?  In this case a solution eigenvector $v^*$ may have exponentially many nonzeros, which is naturally a major obstacle for efficient algorithms or heuristics.  We can circumvent this by asking for statistics about $v^*$ instead.  One option is to instead request samples from a distribution over the labels of the elements of $v^*$, such that the label $l$ is obtained with probability proportional to $(v^*_l)^2$ (or $|v^*_l|^2$ if $H$ is Hermitian and $v^*$ is complex).  This output model still captures Max Cut, since we would obtain some optimal set $S^*$ as a label with probability 1.  Such models are perhaps as not as foreign as they might first appear; for example, Markov Chain Monte Carlo methods are tailored for similar settings.  

Polynomial-time quantum algorithms for linear algebra (see the primer, \cite{dervovic2018quantum}) and machine learning (see the survey, \cite{biamonte2017quantum}) are known in the above kind of model, where the implicitly defined matrices involved are exponentially larger than the number qubits necessary to describe them, and output vectors are only accessible through samples.  However, there are some critical caveats for obtaining exponential quantum advantages in this context \cite{aaronson2015read,aaronson2022much}.  In breakthrough work, quantum-inspired polynomial-time \emph{classical} algorithms of the same flavor have been recently discovered (e.g., \cite{tang2019quantum,chia2020sampling}); however, they rely on a particular kind of classical data access model that may be impractical \cite{cotler2021revisiting}.  Recent empirical demonstrations of quantum advantages are also based on sampling problems \cite{arute2019quantum,madsen2022quantum}.  

\paragraph{\bf The Local Hamiltonian problem.}
Returning to the problem of computing $\lambda_{\max}(H)$ as described above, replacing ``diagonal'' with ``Hermitian'' in the requirements on $H$ takes us from an \NP-complete discrete optimization problem (e.g., Max Cut) to a fundamental \emph{quantum} optimization problem: the \emph{Local Hamiltonian} problem.  Here $H$ is called a \emph{Hamiltonian}, and ``local'' refers to a kind of succinct implicit description of $H$, in the vein of our Max Cut example above.  Physical systems may be described by local Hamiltonians that dictate how they evolve over time, where the eigenvectors of the Hamiltonian correspond to stable states of the system.  In fact nature is constantly trying to solve optimization problems all around us!  Indeed nature strives to heuristically put physical systems in their \emph{ground} states, corresponding to minimum-eigenvalue eigenvectors of the corresponding Hamiltonian.  Consequently, studying ground states of physical systems is a fundamental problem that aids in better understanding and exploiting exotic properties of materials, for example, \cite{minkel2009strange}.

 From a computational perspective, Local Hamiltonian is a cornerstone in understanding the power and limitations of different models of quantum computing, serving a role akin to that of the Boolean Satisfiability problem (SAT) in classical complexity theory.  Local Hamiltonian is complete for the complexity class Quantum Merlin Arthur (\QMA), which contains and is the natural quantum analogue of \NP.  We do not expect polynomial-time quantum algorithms to solve \QMA-hard problems (quantum advantages are more subtle than solving \NP-hard problems \cite{aaronson2016complexity,aaronson2022much}.  Yet, as with \NP-hard problems in the classical regime, aspiring to solve \QMA-hard problems may spark new approaches for heuristic solutions, rigorous approximations, or exact solutions in special cases. 

\paragraph{\bf The Quantum Max Cut problem.}
A desire to help shape the nascent field of quantum approximation algorithms underlaid my foray into QIS.  Sevag Gharibian and I introduced \emph{Quantum Max Cut}, an instance of Local Hamiltonian that is closely related to both the Heisenberg model, a physical model of quantum magnetism, as well as classical Max Cut \cite{gharibian2019almost}.  It turns out that the celebrated Goemans-Williamson SDP-based approximation algorithm for Max Cut \cite{goemans1995improved} can be generalized to give approximation algorithms for Quantum Max Cut \cite{gharibian2019almost}.  The SDP relaxation employed for Max Cut is an instance of the Lasserre/SoS hierarchy, and relaxations for Quantum Max Cut \cite{parekh2021application,parekh2022optimal} may be obtained from a non-commutative\footnote{Max Cut may be cast as $\max_{z_i} \sum_{ij \in E} (1-z_i z_j)/2$ for commutative variables $z_i^2 = 1$, while Quantum Max Cut is $\max_{x_i,y_i,z_i}  \lambda_{\max} (\sum_{ij \in E} (1 - x_i x_j - y_i y_j - z_i z_j)/4)$, for non-commutative variables (i.e., matrices) $x_i^2 = 1$, $y_i^2 =1$, $z_i^2 = 1$ with the additional constraints that variables with different indices commute, while different variables with the same index anti-commute.} counterpart of the Lasserre/SoS hierarchy \cite{pironio2010convergent}. 

As as a canonical constraint-satisfaction and discrete-optimization problem, studying Max Cut has had far-reaching consequences in both computer science \cite{khot2007optimal} and OR \cite{deza1997geometry}, including exponential lower bounds on polyhedral formulations of the Traveling Salesperson problem (via the related correlation polytope) \cite{fiorini2015exponential}.  The goal is for Quantum Max Cut to serve as a testbed for designing approaches to better solve more general Local Hamiltonian problems \cite{parekh2021beating,parekh2022optimal}.  See Section 7 of \cite{hwang2021unique} for an introduction to Quantum Max Cut and Section 3 of \cite{parekh2022optimal} for additional parallels between Max Cut and Quantum Max Cut. 

\paragraph{\bf Challenges.}
Although analogies between Max Cut and Quantum Max Cut have helped direct research into the latter, it largely remains enigmatic.  The Goemans-Williamson $0.878$-approximation for Max Cut \cite{goemans1995improved} is the best possible under the Unique Games Conjecture \cite{khot2007optimal}.  Although an optimal approximation for Quantum Max Cut is known in a special setting \cite{parekh2022optimal}, the currently best-known approximations for the general problem seem far from optimal \cite{hwang2021unique}.  Thus a primary challenge is better approximation algorithms for Quantum Max Cut, as well as more general local Hamiltonians problems. 

Another direction is designing effective and practical heuristics.  I expect OR-influenced heuristics will likely be different and complementary to those currently employed by physicists.  More sophisticated OR-style relaxations, based on bespoke valid inequalities or new types of mathematical-programming hierarchies, are unexplored.  Better understanding the non-commutative Lasserre/SoS hierarchy for Local Hamiltonian is a promising direction, since this may also yield new types of quantum entanglement constraints \cite{parekh2022optimal}.  Finally, Max Cut and Quantum Max Cut are natural unconstrained optimization problems; models, relaxations, and approximations for \emph{constrained} Local Hamiltonian problems are virtually nonexistent.

\subsection{Additional applications}
I mention a few more applications in passing.   $\QIP$ is a model of computation based on quantum interactive proofs, while $\PSPACE$ is the model in which a classical computer is granted polynomial space (but no explicit limit on execution time).  Proofs of the celebrated result that $\QIP = \PSPACE$ rely on the multiplicative weights method for solving SDPs \cite{jain2010qip}.  A recent demonstration of self-concordant barrier functions for quantum relative entropy programs implies more efficient interior-point approaches for such problems \cite{fawzi2022optimal}.  Non-commutative SDP hierarchies can be used to better understand mutually unbiased bases, which have many applications in quantum information and beyond \cite{gribling2021mutually}.  There are a wide variety of further such applications in QIS, and there are likely many more waiting to be discovered.

\subsection{Discerning a quantum advantage}
How will we know if we have witnessed a true and significant quantum advantage?  On the theoretical side, worst-case and asymptotic analysis suggests provable exponential quantum advantages are possible on fault-tolerant quantum computers; however, as previously discussed, problems admitting provable quantum advantages may not have direct classical counterparts, rendering an apples-to-apples comparison difficult.  Even when such comparisons are possible, the big question is whether nature will allow us to engineer scalable quantum computers beyond the near-term noisy intermediate-scale quantum (NISQ) regime \cite{preskill2018quantum}.  

On the practical side, promising empirical benchmarks of current early stage NISQ computers may not be indicative of sustainable advantages into the future.  In addition, recent work points to theoretical limitations of NISQ computing \cite{chen2022complexity,aharonov2022polynomial,kalai2020argument}.  Empirical benchmarks are typically focused on a relatively small or otherwise limited set of instances, and even when quantum computers appear superior, better classical algorithms may be right around the corner.  To mitigate such factors, identifying problems appearing to admit empirical quantum advantages and issuing open challenges in the vein of the DIMACS Implementation Challenges \cite{dimacs2022challenges} may be a fruitful practice.  While this is not a direct QIS application of OR, it is something the latter may share of its culture to benefit the former.

\section{\bf Quantum-inspired Operations Research Advances}
\paragraph{\bf Diversifying instance libraries.}
OR should be proud of its well-defined problem models, curated libraries of problem instances, and systematic benchmarks across ranges of solvers (e.g., \cite{mittelmann2021decision}).  QIS-inspired instances of optimization problems are likely to have a different character than traditional OR instances and would make nice additions to existing instance libraries.  Moreover, solving such instances effectively may drive improvements or new techniques or features in solvers.  Systematic benchmarks and instance collections of quantum-inspired problems may also be of benefit to the QIS community.

\subsection{Quantum algorithms for classical optimization problems}
Quantum advantages are known for a variety of bread-and-butter optimization problems within OR.  Pursuing polynomial-factor quantum speedups for optimization problems including gradient descent \cite{gilyen2019optimizing,kerenidis2020quantum},  linear programming \cite{nannicini2022fast}, second-order cone programming \cite{kerenidis2021quantum}, semidefinite programming \cite{huang2022faster,augustino2021quantum}, and convex programming \cite{chakrabarti2020quantum,van2020convex} has been a fruitful endeavor.  Yet such quantum algorithms do not always improve upon classical counterparts for all the problem parameters, and lower bounds suggest exponential quantum speedups are not possible \cite{garg2021near,chakrabarti2020quantum,van2020convex}.  A ``natural'' or ``practical'' optimization problem admitting a rigorous exponential quantum advantage remains elusive.  

In place of a speedup, we may consider how well a quantum algorithm might \emph{approximate} a problem relative to classical algorithms.  The Quantum Approximate Optimization Algorithm (QAOA) \cite{farhi2014quantum} is a quantum-algorithmic framework for formulating and solving discrete optimization problems (see the thesis, \cite{hadfield2018quantum}).  QAOA resembles mathematical programming in that discrete optimization problems may be relatively easily expressed in the framework, and the overall efficacy of the algorithm is often highly dependent on the particular formulation employed.  QAOA has garnered considerable QIS community interest, and it is a natural candidate for implementation on NISQ systems.  Moreover, QAOA is a parameterized algorithm that lends itself to a natural hybrid quantum-classical loop, wherein a classical optimization routine is used to obtain parameter values by leveraging a quantum computer to execute QAOA.  The performance of QAOA is assessed on the current parameter values, informing subsequent parameter updates.  It is not currently known if there is a problem where a polynomial-time execution of QAOA on a quantum computer is able to yield a provably better approximation than possible classically.  Some theoretical limitations of QAOA are known \cite{anshu2022concentration,chou2022limitations,marwaha2022bounds,bravyi2020obstacles}.  Even if QAOA is unable to provide an advantage on classical problems, QAOA might be able to achieve better approximations than possible classically for \emph{quantum} optimization problems, such as Quantum Max Cut.  Such directions are understudied \cite{anshu2020beyond,anshu2021improved}.  Better understanding the power and limitations of QAOA, applying it to broader problem classes, and devising QAOA-inspired classical algorithms remain challenges.  

\subsection{New types of problems inspired by quantum physics}
Beyond helping to solve quantum optimization problems, I urge operations researchers to allow quantum problems to inspire new types of classical ones.  Let me supply an example.  Operations researchers are well aware that our models frequently fail to capture the nuances of the underlying practical problems we endeavor to solve.  For this reason and others, a diverse set of near-optimal solutions may be preferable to a single optimal solution.  

Consider the following version of Max Cut that seeks to promote diverse solutions: 
\begin{equation*}
\max_{\mathbf{x}} \ \mathbb{E} [c(\mathbf{x})] + \sum_{x,y \in \{0,1\}^n} \sqrt{\text{Pr}[\mathbf{x} = x] \text{Pr}[\mathbf{x} = y]} h(x,y), 
\end{equation*}    
where $\mathbf{x} \in \{0,1\}^n$ is a random variable, $c(x)$ is the cost of the cut induced by $x$, and $\{0,1\} \ni h(x,y) = 1$ if and only if $x_i \not= y_i$ for exactly one position $i$.  The above problem then seeks to find a distribution over cuts, represented by $\mathbf{x}$, that is incentivized for both having a large expected cut value as well as having support on pairs of cuts that differ in exactly one vertex.  Here $h(x,y)$ is meant to measure diversity between $x$ and $y$, and other options are possible.  
%However, the above $h(x,y)$ is perhaps interesting because it encourages a somewhat hierarchical notion of diversity.  There is no objective term for $x$ and $y$ that differ in more than one position, but if $h(x,y)=1$, $h(y,z)=1$, both terms are credited in the objective. 

The notion of diversity above arises naturally as what is known as a \emph{transverse-field} term in the quantum Ising model studied in statistical mechanics.  We could think of the above problem as a transverse-field Max Cut problem, and it is actually equivalent to the well-studied transverse-field Ising model, which is a Local Hamiltonian problem.  Fresh insights into the problem may have an impact on the physical side of things, as I already argued for more general Local Hamiltonian problems.  However, here I would like to emphasize a complementary story.  We may readily adapt $\textbf{x}$ and $c(x)$ above to derive transverse-field versions of other familiar optimization problems, where we likewise seek distributions over diverse solutions.  In this case insights gleaned from a physical perspective might suggest new avenues for these and related optimization problems.  The challenge is then to more broadly frame physical models and solution techniques in a way that might impact OR.      
 
\section{Suggestions for Engaging QIS} 
QIS is a diverse and multi-disciplinary field collecting physicists, chemists, engineers, computer scientists, mathematicians and increasingly, operations researchers.  Here the notion of an ``outsider'' is perhaps diminished relative to more homogeneous fields.  Reach out to your friendly neighborhood quantum information scientist with questions or for guidance.  Seek out OR colleagues who have started dabbling in QIS.  Members of the INFORMS ICS Working Group on Quantum Computing and others have been active in organizing workshops bridging OR and QIS -- be on the lookout for such opportunities.   

\paragraph{\bf Suggested reading.}
I have explicitly called out references to surveys, primers, tutorials, reviews, and theses above as hints for further reading.  As far as textbooks go, Nielsen and Chuang is the standard introduction to quantum computing and quantum information for good reason \cite{nielsen2010quantum}.  Several quantum information scientists maintain web pages with pointers for learning QIS topics (e.g., \cite{harrow2022learn}).

I want to stress that you do not have to necessarily understand quantum physics to have an impact.  Working in interdisciplinary teams or settling into a comfortable mathematical formalism are some ways to mitigate a proper physics background.  If your regular time commitments are at odds with cover-to-cover reading of new material, of course you should not feel ashamed to focus on the bits and pieces that interest you most, with the hope that you will eventually be able to fill in gaps as necessary or time permits.  If you would like to roll up your sleeves and learn interactively, quantum programming tutorials such as \cite{matthews2021get} might be a good place to start. 

\section*{Concluding Remarks} Improved capabilities for solving optimization problems arising in quantum physics can help us better understand how nature works at a fundamental level.  Both practical approaches for solving specific kinds of problems as well as furthering theoretical foundations are valuable.  Historically, a stronger command of nature has fueled transformative impacts on civilization and society. 

\section*{Acknowledgements}
This material is based upon work supported by the U.S. Department of Energy, Office of Science, Office of Advanced Scientific Computing Research, Accelerated Research in Quantum Computing, Fundamental Algorithmic Research for Quantum Computing (FAR-QC).

This article has been authored by an employee of National Technology \& Engineering Solutions of Sandia, LLC under Contract No. DE-NA0003525 with the U.S. Department of Energy (DOE). The employee owns all right, title and interest in and to the article and is solely responsible for its contents. The United States Government retains and the publisher, by accepting the article for publication, acknowledges that the United States Government retains a non-exclusive, paid-up, irrevocable, world-wide license to publish or reproduce the published form of this article or allow others to do so, for United States Government purposes. The DOE will provide public access to these results of federally sponsored research in accordance with the DOE Public Access Plan \url{https://www.energy.gov/downloads/doe-public-access-plan}.

\bibliographystyle{alpha}
\bibliography{QIS-OR}

\end{document}